# Interpretation of "non-local" experiments using disentanglement.

By

B. C. Sanctuary[*], S. Pressé[†], T. D. Lazzara, E. J. Henderson, and R. F. Hourani

Department of Chemistry
McGill University
801 Sherbrooke Street W
Montreal, Quebec
H3A 2K6, Canada

**Abstract**

It is shown that a number of experiments designed to use entangled photon pairs in order to demonstrate the viability of quantum "teleportation" can, in fact, also be understood using disentanglement. Whether entangled or not, using an ensemble approach, the experiments can be explained without any non-local communication between Alice's photon and Bob's photon. Moreover, it is emphasized that entanglement maintains a symmetry property between the two photons that is absent in disentanglement, the symmetry being parity due to phase coherence.

**Introduction**

Recently[1] it was shown that after separation an EPR-entangled pair carries two types of correlation. One arises from the quantum interference terms and is identified with conservation of parity, while the other arises from the "classical" terms and is identified with conservation of angular momentum. When both are present, the correlation function for a pair of photons has the form $E(\mathbf{a},\mathbf{b}) = -\cos 2\theta_{ab}$ where the angle is between the two unit vectors, $\mathbf{a}$ and $\mathbf{b}$, that orient the two polarizers in EPR-type experiments. When parity is not conserved the correlation reduces to $E(\mathbf{a},\mathbf{b}) = -\tfrac{1}{2}\cos 2\theta_{ab}$. The former can lead to violations of Bell's inequalities while the latter always satisfies them. Since the functional form of the correlations is the same, $\cos 2\theta_{ab}$, many experimenters[2,3,4,5,6] have assumed that the separated photons are entangled whereas in fact there is evidence they might not remain so.

It has been argued[7] that whether entangled or not, once separated and beyond the range of any interaction photons from an EPR pair cannot in any way influence one another. Using a description that treats the singlet state as a statistical ensemble of spins with different quantization axes, it follows that[8] "teleportation" experimental results can be explained without recourse to non-locality. In this paper, we discuss this alternate interpretation and these show that the experiments are in fact selecting states from an ensemble of those possible.

---

[*] Correspondence should be addressed to bryan.sanctuary@mcgill.ca
[†] Present address: Department of Chemistry, MIT, Boston, MA

Although this paper does not suggest that entangled states cannot be maintained after separation[9], it is shown that a number of experiments can be explained effectively using disentanglement[2,3,4,5,6]. The study of disentangled states has merit since long distance correlation is predicted without recourse to non-locality. Confirmation of this would give experimental justification for interpreting the wave function as describing the state as statistical ensemble rather than a single quantum entity, such as a photon.

**Double coincidence Aspect-type experiments**

It has been pointed out[1] that the Aspect[10] experiments are not sensitive to the overall pre-factor that distinguishes between entangled and disentangled EPR pairs. The correlation function, $E(\mathbf{a},\mathbf{b})$ can be related to the four double coincident probabilities by

$$E(\mathbf{a},\mathbf{b}) = P_{++}(\mathbf{a},\mathbf{b}) - P_{+-}(\mathbf{a},\mathbf{b}) - P_{-+}(\mathbf{a},\mathbf{b}) + P_{--}(\mathbf{a},\mathbf{b}) \qquad (1.1)$$

These probabilities are determined by counting coincident events but because of losses of various kinds, the probabilities are written with an empirical parameter call the visibility, $V$,

$$P_{+-}(\mathbf{a},\mathbf{b}) = \frac{1}{4}(1 + V\cos 2\theta_{ab}) \qquad (1.2)$$

If the photons are entangled then the visibility is $V=1$. If the photons are completely disentangled the visibility is $V=1/2$. The other probabilities occurring in Eq.(1.1) are similarly defined[1] leading to a correlation of

$$E(\mathbf{a},\mathbf{b}) = -V\cos 2\theta_{ab} \qquad (1.3)$$

Since the Aspect[10] experiments are not sensitive to the visibility, it is questioned as to whether such experiments can demonstrate that QM is a non-local theory.

In the repeats of the Aspect experiments Weihs *et al*[11] agree with the Aspect results. Again, however, by retaining only coincidence counting, and normalizing the probabilities with respect to the total number of such counts, these experiments also do not establish the value of the visibility as defined in Eq.(1.2). Therefore, the Aspect[10] and Weihs *et al.*[11] experiments cannot confirm whether entanglement is maintained as the particles separate. However, two points are consistent with the conclusion that the photons are disentangled.

First, if the photons are entangled then random coincidences are not predicted. In fact experimentally random coincidences are observed but are treated as errors and removed[2,11]. Analysis of these random coincidences gives evidence for the presence of disentangled photons[1]. Second, the maximum theoretical detection rate predicted from entangled photon pairs is 25 % while from disentanglement it is predicted to be[1] 6.25 %. Weihs *et al.*[11] report a detection rate of 5 %. In the absence of further experimental evidence, it is not conclusive in these experiments that entanglement is maintained up to the filters even though they may have been initially entangled.

The 10 km experiment of Gisin *et al.*[3], display an uncorrected visibility of $V$=0.46 and this is consistent with the results from disentanglement. This reduced visibility is explained[3] by being due to the high number of accidental coincidences which are subtracted, following the work of Brendel *et al.*[12]. This increases the visibility to 87%. However, these experiments clearly show an uncorrected visibility of about 50%, consistent with disentanglement, again suggesting that the photons do not retain entanglement after leaving the source.

Quantum "teleportation" experiments are considered to give definitive evidence for the non-local nature of quantum mechanics. These experiments are extensions of the Aspect-type experiments from double coincidences to 3 and 4 coincidence experiments.

**The experiments of Gisin *et al.*[4]**

In a recent paper by Gisin *et al.*[4], a triple coincidence "teleportation" experiment was performed where the state measured by Charlie is the singlet Bell state, $\left|\Psi_{12}^{-}\right\rangle$, while at Bob's location the detected state is given by

$$\left|\Psi_{3}(\beta)\right\rangle = \left(a_{o}\left|+\right\rangle_{\hat{z}}^{3} + e^{i\beta}a_{1}\left|-\right\rangle_{\hat{z}}^{3}\right) \tag{2.1}$$

The phase angle β is an adjustable parameter and $a_o$ and $a_1$ can be chosen to be 0, 1 or $\frac{1}{\sqrt{2}}$ such that $|a_0|^2 + |a_1|^2 = 1$. The measured state is $\left|\Phi_{123}\right\rangle = \left|\Psi_{12}^{-}\right\rangle\left|\Psi_{3}(\beta)\right\rangle$ which[7] leads to the expectation value as a function β. Assuming entanglement the result is,

$$\left\langle\Phi_{123}\left|\rho^{1}\rho^{23}\right|\Phi_{123}\right\rangle = \frac{1}{8}(1-\cos\beta) \tag{2.2}$$

whereas that from disentanglement is,

$$\overline{\left\langle\Phi_{123}\left|\rho^{1}\rho_{Disentanglement}^{23}\right|\Phi_{123}\right\rangle} = \frac{1}{8}\left(1-\frac{1}{2}\cos\beta\right) \tag{2.3}$$

The bar denotes ensemble averaging[7]. In this experiment, $\rho^{23} = \left|\Phi_{23}^{+}\right\rangle\left\langle\Phi_{23}^{+}\right|$ in Eq.(2.2) is one of the four Bell states.

Equations (2.2) and (2.3) are plotted and compared with the experimental results in Figure 1. The results from disentanglement agree with experiment as well as those from entanglement. One difference is that the entanglement results predict no offset (or base line) while experimentally one is clearly present. This baseline is predicted using disentanglement. Moreover, the ratio of the offset to the maximum intensity is 1:3 and this is within the uncertainty of the experiment. The detection rate is small, being roughly 20 coincidences in a time of 500 seconds, far below the theoretical maximum of 25% predicted from entanglement. This low detection rate is consistent with the properties of Alice's photon matching only a few of those produced in the disentangled ensemble. Further discussion of the low detection rate from disentanglement is deferred to the discussion.

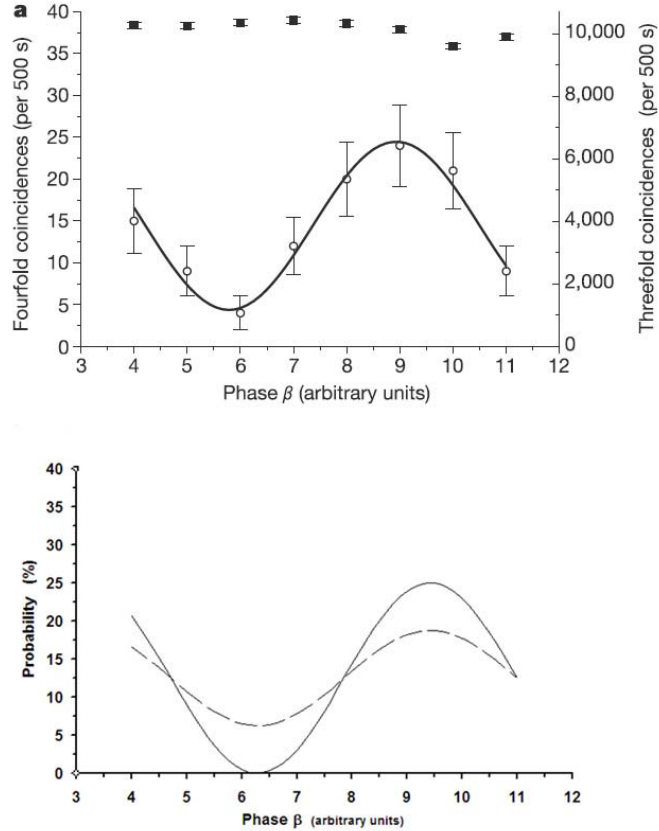

Figure 1 (upper) The experimental results of Gisin et al.[4] (lower) A plot of Eq.(2.3) obtained from disentanglement (dashed curve) and of Eq.(2.2) from entanglement (solid curve). Experiment, figure (upper), shows both triple (filled squares) and fourfold (open circles) coincidences. Since the triple coincidences plot contains accidental coincidences, and these have been removed in the fourfold coincidence experiments, theory, (figure (1b)), is directly comparable to the fourfold coincidence plot.

**The experiment of Zeilinger et al.[5]**

In a similar experiment[5] in 1997, "teleportation" results using the Bell state $\left|\Psi_{12}^{-}\right\rangle$ were presented in which Alice's photon is chosen to be initially polarized either as +45° or -45° and the source generates entangled states in the singlet state. A polarizing beam splitter at Bob's location deflects the photons towards two detectors, sensitive to photon 3 having polarization of +45° and -45°. If Alice's photon is initially polarized at +45°, then "teleportation" is considered to have occurred if Bob's +45° detector records a coincidence while the -45° returns no response. The opposite results are expected if Alice's initial state is -45°. If the measured state is represented as $|\Phi_{123}\rangle = |\Psi^-_{12}\Psi_3(\pm 45°)\rangle$ then the theoretical predictions from entanglement, give a value of ¼ for Alice's state being $\rho^1 = \rho^1(+45°)$ and a value of zero for $\rho^1 = \rho^1(-45°)$. The theoretical results from disentanglement are

$$\overline{\langle \Psi^-_{12}\Psi_3(\pm 45°)|\rho^1(\pm 45°)\rho^{23}_{disentangled}|\Psi^-_{12}\Psi_3(\pm 45°)\rangle} = \frac{1}{8}\left(1+\overline{2\cos^2\theta\sin^2\theta}\right) = \frac{3}{16} \quad (3.1)$$

and

$$\overline{\langle \Psi^-_{12}\Psi_3(\pm 45°)|\rho^1(\mp 45°)\rho^{23}_{disentangled}|\Psi^-_{12}\Psi_3(\pm 45°)\rangle} = \frac{1}{8}\left(1-\overline{2\cos^2\theta\sin^2\theta}\right) = \frac{1}{16} \quad (3.2)$$

Here the cylindrical ensemble average is over the angle $\theta$ and gives $\overline{\theta} = 45°$. Table 1 summarizes the results and can be compared to the experiment in Figure 2.

| Photon 1 | Photon 3 | Entanglement | Disentanglement | Experimental |
|---|---|---|---|---|
| ±45° | ±45° | 100 % | 100 % | 100 % |
| ±45° | ∓45° | 0.0 % | 33.3 % | ≈ 15 – 40 % |

Table 1. Calculations from entanglement and disentanglement giving the relative intensities of the predicted results..

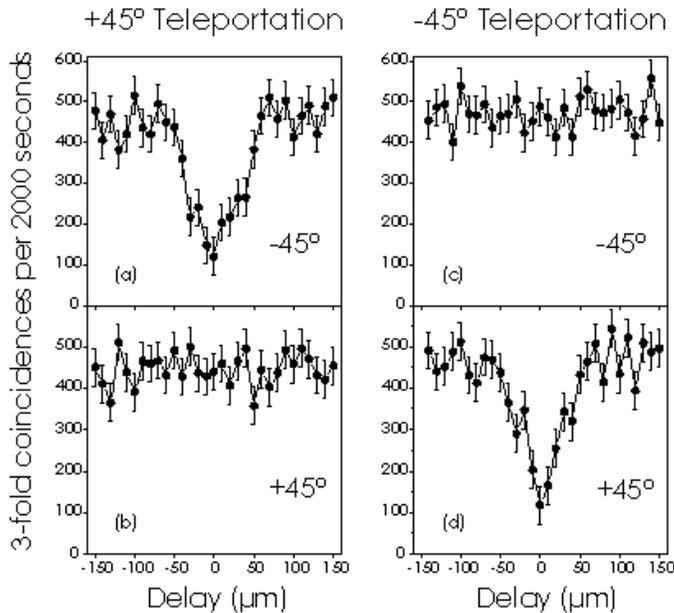

Figure 2. Experimental results from the Zeilinger et al.[5] "teleportation" experiment showing that the dip in intensity does not go to zero as predicted from entanglement, and is consistent with disentanglement.

Entanglement and disentanglement both predict a dip in the triple coincidence rates, when the state of Bob's photon does not match that of Alice. Disentanglement predicts that the intensity does not go to zero in contrast to entanglement. Moreover the ratio, 1:3, obtained from disentanglement, is well within experimental error. Therefore the results appear better described using disentanglement rather than entanglement. The detection rate is also low.

**The experiments of Kim *et al.*[6]**

Recently Kim *et al.*[6] used non-linear crystals in experiments that allow for all four of the Bell states to be measured at Alice's location using detectors labeled $D_4^I$ through $D_4^{IV}$. Such crystals have the property of changing the helicity states of the photons 1 and 2. More specifically, type-I SFG crystals cause the following transformations: $|++\rangle \to |--\rangle$ and $|--\rangle \to |++\rangle$. Likewise, type-II SFG crystals cause states to change as follows: $|+-\rangle \to |-+\rangle$ and $|-+\rangle \to |+-\rangle$. In the disentanglement procedure, the axis of quantization is oriented by angles $\theta, \phi$. As a photon passes through type-I crystals, the azimuthal angles are changed from $\phi \to \phi + \pi$. In the experiments, detector responses at Alice's location are combined with detector responses at Bob's location in order to extract coincidences between detectors $D_4^I$ through $D_4^{IV}$ and Bob's detector $D_3$. An analyzer is placed in front of Bob's detector, so that the coincidences are measured as a function of the angle $\varphi$, of this analyzer with range $0 < \varphi \leq 2\pi$. If Alice's photon is initially in state, $|\psi_1\rangle = \begin{pmatrix} \alpha \\ \beta \end{pmatrix}$ then detector $D_4^I$ detects state $\begin{pmatrix} \alpha \\ -\beta \end{pmatrix}$ while detector $D_4^{II}$ detects state $\begin{pmatrix} \alpha \\ \beta \end{pmatrix}$. Alice's photon is +45° polarized, i.e. $\alpha = \beta = \frac{1}{\sqrt{2}}$. For the coincident $D_4^I - D^3$ and $D_4^{II} - D^3$ projection measurements, the observable states are respectively,

$$|\Phi_{123}^I(\varphi)\rangle = \frac{1}{\sqrt{2}} \begin{pmatrix} 1 \\ 0 \\ 0 \\ 1 \end{pmatrix} \begin{pmatrix} \cos\varphi \\ \sin\varphi \end{pmatrix}; \text{ and } |\Phi_{123}^{II}(\varphi)\rangle = \frac{1}{\sqrt{2}} \begin{pmatrix} -1 \\ 0 \\ 0 \\ 1 \end{pmatrix} \begin{pmatrix} \cos\varphi \\ \sin\varphi \end{pmatrix} \qquad (3.3)$$

Both *l=I*, and *l=II*, lead to the results from entanglement,

$$\langle \Phi_{123}(\phi) | \rho^1 \rho_{EPR}^{23} | \Phi_{123}(\phi) \rangle = \frac{1}{8}(1 \pm 2\cos\varphi\sin\varphi) \qquad (3.4)$$

where the "+" sign is for detector coincidence *l*=1, $(D_4^I - D^3)$ and the "-" sign is for detector coincidence *l*=2, $(D_4^{II} - D^3)$. Likewise, the results from disentanglement after ensemble averaging are,

$$\overline{\langle \Phi_{123}(\phi) | \rho^1 \rho^{12}_{\hat{P},\text{disentangled}} | \Phi_{123}(\phi) \rangle} = \frac{1}{8}(1 \pm \cos\varphi \sin\varphi) \qquad (3.5)$$

Equations (3.4) and (3.5) are plotted in Figure 3 along with the experimental results[6]. The experimental curves display a offset that is predicted from disentanglement, which is not present when entangled states are assumed. Moreover, although the experimental ratio of offset intensity to oscillation peak intensity is about 1:5, the error bars are large, so that the disentanglement curve, 1:3, falls within experimental error.

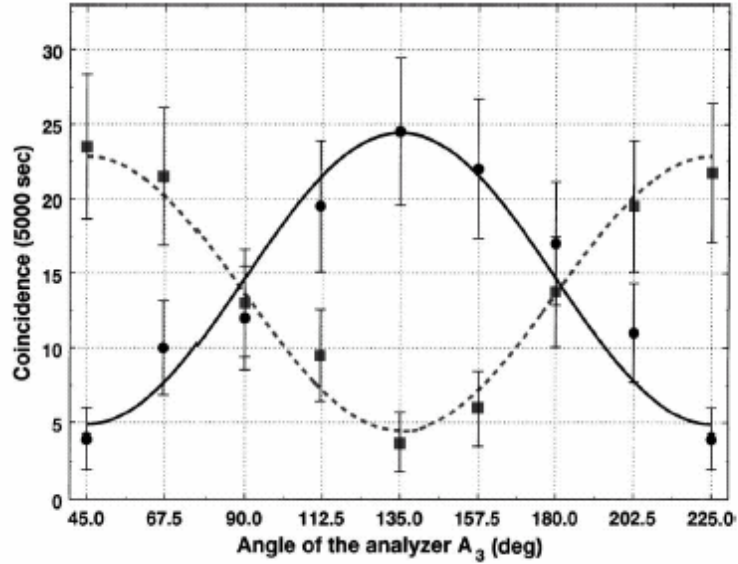

(a)

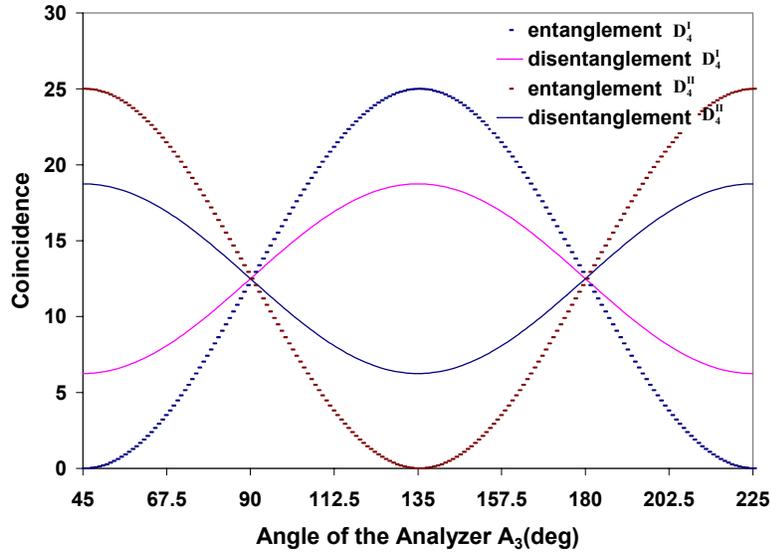

(b)

Figure 3 (a) the experimental results of Kim *et al.*[6] plotted as a function of analyzer angle. (b) Plot of Eq.(3.4) from entanglement (dotted cureves) and a plot of Eq. (3.5) from disentanglement (solid curves). In the figures, two different detector coincidences are shown: Those curves that maximize the intensity at 135º, are plots of $(D_4^I - D^3)$ while those that minimize the intensity at 135º are plots of $(D_4^{II} - D^3)$.

**Discussion**

In this paper, we have calculated, assuming the separated photons are either entangled or disentangled, various theoretical expectation values for some coincidence experiments designed to study quantum "teleportation". It is shown that disentanglement gives agreement that is at least as good as those from entanglement. It can therefore be concluded that the photons, which are experimentally generated from the source, are either not entangled, or disentangle somewhat, or completely, before detection. As further evidence of this, low detection rates are consistent with disentanglement. The detection rates are typically of the order of 0.01%, much less than the maximum predicted theoretically of the 25%.

The process of disentanglement generates random phases between the disentangled photons. Denoting the phases by $\phi_2$ and $\phi_3$, only a small fraction of photons with approximately equal phases can lead to successful quantum state matching between photons from the ensemble. To obtain a quantitative measure of how close the phases must be, define the phase difference as

$$\delta\phi = \phi_3 - \phi_2 \qquad (4.1)$$

Assuming a detection rate of 0.01 shows the phase matching must obey approximately,

$$|\delta\phi| < 0.58° \qquad (4.2)$$

If the photons remain entangled, then the phases always match perfectly and $\delta\phi = 0$ for all EPR pairs which belong to an ensemble containing only four states. Hence disentanglement accounts for the low detection rates.

The treatment here is based upon the singlet state being represented by an ensemble of EPR pairs, all with different quantization axes. Disentanglement reveals these axes so that Alice's photon can only form Bell states with a specific sub-ensemble of the singlet state.

**Acknowledgement:**

*This work was supported by a research grant from the Natural Sciences and Research Council of Canada (NSERC).*